\documentclass[12pt]{article}

\usepackage{amsfonts}
\usepackage{amssymb}
\usepackage{amscd}

\def\D{\hbox{D\kern-.73em\raise.25ex\hbox{-}\raise-.25ex\hbox{ }}}
 \def\d{\hbox{d\kern-.33em\raise.75ex\hbox{-}\raise-.75ex\hbox{}}}

\def\GGG{\frak G }
\def\gr3{\GGG\,(\SSS_3)}

\def\gr2{\GGG\,(\SSS_2)}

\def\SSS{\frak S}

\def\ed{\end{document}}
\def\beq{\begin{equation}}
\def\eeq{\end{equation}}
\def\bea{\begin{eqnarray}}
\def\eea{\end{eqnarray}}
\def\ba{\begin{array}}
\def\ea{\end{array}}
\def\bi{\begin{itemize}}
\def\ei{\end{itemize}}

\def\nn{\nonumber}

\newcommand{\bp}{\noindent\begin{minipage}[c]}
\newcommand{\ep}{\end{minipage}}

\begin{document}
\title{\bf Zeta Strings }

\author{Branko Dragovich\thanks{\,e-mail
address: dragovich@phy.bg.ac.yu} \\ {}\\
\it{Institute of Physics}\\ \it{Pregrevica 118, P.O. Box 57, 11001
Belgrade, Serbia}}

\date {~}
\maketitle
\begin{abstract}
We introduce  nonlinear scalar field models for open and
open-closed strings with spacetime derivatives encoded in the
operator valued Riemann zeta function. The corresponding two
Lagrangians are derived in an adelic approach starting from the
exact Lagrangians for effective fields of $p$-adic tachyon
strings. As a result tachyons are absent in these models. These
new strings we propose to call zeta strings. Some basic classical
properties of the zeta strings are obtained and presented in this
paper.
\end{abstract}

\section{Introduction}

There are just twenty years from the publication of  the first
paper on the $p$-adic string  \cite{volovich1}. So far $p$-adic
structures have been observed not only in string theory but also
in many other models of modern mathematical physics (for a review
of the early days developments, see e.g. \cite{freund1},
\cite{volovich2}).

One of the greatest achievements in $p$-adic string theory is an
effective field description of scalar open and closed $p$-adic
strings \cite{freund2}, \cite{frampton1}. The corresponding
Lagrangians are very simple and exact. They describe not only
four-point scattering amplitudes but also all higher ones at the
tree-level.

This $p$-adic string theory  has been significantly  pushed
forward when was shown \cite{sen} that it describes tachyon
condensation and brane descent relations simpler than by ordinary
bosonic strings. After that success, many aspects of $p$-adic
string dynamics have been investigated and compared with dynamics
of ordinary strings (see, e.g. \cite{many} and references
therein). Noncommutative deformation of $p$-adic string
world-sheet with a constant B-field was  investigated in
\cite{ghoshal-grange} (on $p$-adic noncommutativity see also
\cite{dragovich1}). A systematic mathematical study of spatially
homogeneous solutions of the nonlinear equation of motion was
performed in \cite{vladimirov}. Some possible cosmological
implications of $p$-adic string theory have been also investigated
\cite{arefeva-barnaby}. It was recently proposed \cite{ghoshal}
that $p$-adic string theories provide lattice discretization to
the world-sheet of ordinary strings. As a result of these
developments many nontrivial features of ordinary string theory
have been reproduced from the $p$-adic effective action.

Adelic approach to the string scattering amplitudes is a very
useful way to connect $p$-adic and ordinary counterparts (see
\cite{freund1} as a review). Moreover, it eliminates unwanted
prime number parameter $p$ contained in $p$-adic amplitudes and
also cures the problem of $p$-adic causality violation. Adelic
generalization of quantum mechanics was also successfully
formulated, and was found a connection between adelic vacuum state
of the harmonic oscillator and the Riemann zeta function
\cite{dragovich2}. Recently, an interesting approach  toward
foundation of a field theory and cosmology based on the Riemann
zeta function was proposed in \cite{volovich3}. An adelic approach
with the Riemann zeta function is one of the motivations for this
paper.

The present paper is a result of the attempt to integrate all
$p$-adic effective field actions into one bosonic field theory. In
the next two sections we explore the cases of open and open-closed
bosonic strings.

\section{Open scalar zeta string}

The exact tree-level Lagrangian for effective scalar field
$\varphi$ which describes open $p$-adic string tachyon is

\beq {\cal L}_p = \frac{1}{g^2}\, \frac{p^2}{p-1} \Big[
-\frac{1}{2}\, \varphi \, p^{-\frac{\Box}{2}} \, \varphi  +
\frac{1}{p+1}\, \varphi^{p+1} \Big]\,,  \label{1.1} \eeq where $p$
 is any prime number, $\Box = - \partial_t^2  + \nabla^2$ is the
$D$-dimensional d'Alembertian and we adopt metric with signature
$(- \, + \, ...\, +)$. An infinite number of spacetime derivatives
follows from the expansion
$$
p^{-\frac{\Box}{2}} = \exp{\Big( - \frac{1}{2} \ln{p}\, \Box \Big)}
= \sum_{k \geq 0} \, \Big(-\frac{\ln p}{2} \Big)^k \, \frac{1}{k
!}\, \Box^k \,.
$$
The equation of motion for (\ref{1.1}) is

\beq p^{-\frac{\Box}{2}}\, \varphi = \varphi^p \,,   \label{1.2}
\eeq and its properties have been studied by many authors (see
e.g. \cite{vladimirov} and references therein).

It is worth noting that prime number $p$ in (\ref{1.1}) and
(\ref{1.2}) can be replaced by  natural number $n \geq 2$ and such
expressions  also make sense. Moreover,  when $p = 1 + \varepsilon
\to 1$ there is the limit of (\ref{1.1}) which is related to the
ordinary bosonic string in the boundary string field theory
\cite{gerasimov}.

Now we want to introduce a model which incorporates all the above
$p$-adic string Lagrangians in a restricted adelic way. To this
end, let us take the sum of the above Lagrangians ${\cal L}_n$
(\ref{1.1}) in the form

\bea L &=& \sum_{n\geq 1} C_n\, {\cal L}_n  = \sum_{n\geq 1}
\frac{n-1}{n^2} \, {\cal L}_n \nn \\ &=& \frac{1}{g^2} \Big[
-\frac{1}{2}\, \phi \, \sum_{n\geq 1} n^{-\frac{\Box}{2}} \, \phi
+ \sum_{n\geq 1} \frac{1}{n + 1} \, \phi^{n+1} \Big]\,,
\label{1.3} \eea where  coefficients $C_n = \frac{n-1}{n^2}$ are
inverses of those within ${\cal L}_n$.   Note that this choice is
formally equivalent to the following one: $\frac{1}{g^2}\,
\frac{n^2}{n - 1} \, \to \, \frac{1}{g^2}$ and $C_n = 1$, but it
seems to be less natural. Thus we retain the string coupling
constant $g$.   To emphasize that Lagrangian (\ref{1.3}) describes
a new field, which takes into account all $p$-adic fields, we
introduced notation $\phi$ instead of $\varphi$. The term $C_1 \,
{\mathcal L}_1 = 0$, because of its two equal parts of opposite
sign, but these parts give contribution to  kinetic and potential
terms of the total Lagrangian $L$.

According to the famous Euler product formula one can write
$$ \sum_{n\geq 1} n^{-\frac{\Box}{2}} = \prod_p \frac{1}{ 1 -
p^{-\frac{\Box}{2}}}\,. $$  Recall that the Riemann zeta function
is defined as

\beq  \zeta (s) = \sum_{n\geq 1} \frac{1}{n^{s}} = \prod_p
\frac{1}{ 1 - p^{- s}}\,, \quad s = \sigma + i \tau \,, \quad
\sigma >1\,. \label{1.4} \eeq Employing usual expansion for the
logarithmic function and definition (\ref{1.4}) we can rewrite
(\ref{1.3}) in the form

\beq L = -\frac{1}{g^2} \Big[ \frac{1}{2}\, \phi \,
\zeta\Big({\frac{\Box}{2}}\Big) \, \phi + \phi +  \ln(1 -\phi)
\Big]\,, \label{1.5} \eeq where $ |\phi| < 1 \,.$

$\zeta\Big({\frac{\Box}{2}}\Big)$ acts as  pseudodifferential
operator in the following way (see also \cite{volovich3}): \beq
\label{1.6} \zeta\Big({\frac{\Box}{2}}\Big)\, \phi (x) =
\frac{1}{(2\pi)^D}\, \int e^{ ixk}\, \zeta\Big(-\frac{k^2}{2}\Big)\,
\tilde{\phi}(k)\,dk \,, \quad -k^2 = k_0^2 -\overrightarrow{k}^2  >
2 + \varepsilon\,,
 \eeq
where  $ \tilde{\phi}(k) =\int e^{(-  i kx)} \,\phi (x)\, dx$ is the
Fourier transform of $\phi (x)$. The region of integration in
(\ref{1.6}) is $-k^2 = k_0^2 -\overrightarrow{k}^2  > 2 +
\varepsilon$, where $\varepsilon$ is an arbitrary small positive
number, and it follows from the definition of  zeta function
(\ref{1.4}). Here and in the sequel, it is understood that this zeta
function depends also on $\varepsilon$. The usual tachyon with mass
$ m^2 = -k^2 = - 2$ is absent in this theory and the energy of this
new string is bounded from below by $k_0^2 = 2$ in the string mass
scale.

 Dynamics of this field $\phi$ is
encoded in the (pseudo)differential form of the Riemann zeta
function. When the d'Alembertian is an argument of the Riemann
zeta function we shall call such string a zeta string.
Consequently, the above $\phi$  is an open scalar zeta string.

The equation of motion for the zeta string $\phi$ is

\beq \zeta\Big({\frac{\Box}{2}}\Big) \, \phi =
\frac{1}{(2\pi)^D}\, \int_{ k_0^2 -\overrightarrow{k}^2  > \, 2
+\varepsilon} e^{ ixk}\, \zeta\Big(-\frac{k^2}{2}\Big)\,
\tilde{\phi}(k)\,dk  = \frac{\phi}{1 - \phi} \label{1.7} \eeq
which has an evident solution $\phi = 0$.

The above zeta string potential is given by

\beq V (\phi) = \frac{1}{g^2}\, [\phi + \ln (1 -\phi)] = -
\frac{1}{g^2}\, \sum_{n\geq 2} \frac{\phi^n}{n}\,, \label{1.8}\eeq
where $V (\phi) \leq 0$ for $-1 < \phi < 1$: it increases from $V
(\phi \to -1) = - \frac{0, 31}{g^2}$ to the maximum $V (\phi = 0)
= 0$ and then $V(\phi)$ decreases so that $V (\phi) \to -\infty$
as $\phi \to +1$.

For the case of time dependent spatially homogeneous solutions one
has to consider the equation of motion

\beq \zeta \Big({\frac{- \partial_t^2}{2}}\Big) \, \phi (t) =
\frac{1}{(2\pi)}\, \int_{|k_0|> \, \sqrt{2} + \varepsilon} e^{-
ik_0 t}\, \zeta\Big(\frac{k_0^2}{2}\Big)\, \tilde{\phi}(k_0)\,dk_0
= \frac{\phi (t)} {1 - \phi (t)} \,. \label{1.9} \eeq

In the weak field approximation $(|\phi (t)|\ll 1)$ the above
expression $\phi/(1 - \phi) \thickapprox \phi$ and (\ref{1.9})
becomes a linear equation  which can be written in the form

\beq \int_{\mathbb R}\, e^{- i k_0 t} \,
\Big[\zeta\Big(\frac{k_0^2}{2}\Big)\, \, \theta (|k_0| - \sqrt{2} -
\varepsilon)  - 1\Big]\, \tilde{\phi}(k_0) \, dk_0  = 0 \,,
\label{1.9a} \eeq where $\theta$ is the Heaviside function. Since
$\zeta\Big(\frac{k_0^2}{2}\Big) > 1$ when $|k_0| > \sqrt{2}$ the
equation (\ref{1.9a}) has solution only for $\tilde{\phi}(k_0) = 0$.
This also means the absence of mass.

\section{Open and closed scalar zeta strings}

The exact Lagrangian for the coupled open and closed $p$-adic
tachyons  is presented in \cite{freund1} and it reads

\bea  {\mathcal L}_p' &=& \frac{1}{g^2}\, \frac{p^2}{p-1} \Big[ -
\frac{1}{2} \varphi\, p^{-\frac{\Box}{2}}\, \varphi +
\frac{1}{p+1}\, \psi^{\frac{p(p-1)}{2}}\, (\varphi^{p+1} - 1)
\Big]\nn \\ &+& \frac{1}{h^2}\, \frac{p^4}{p^2 - 1} \Big[-
\frac{1}{2} \, \psi\, p^{-\frac{\Box}{4}}\, \psi + \frac{1}{p^2
+1}\, \psi^{p^2 + 1}\Big]\nn \\   &=& \frac{1}{g^2}\,\, {\mathcal
M}_p (\varphi, \psi) +  \frac{1}{h^2} \,\, {\mathcal N}_p (\psi)
\,, \label{1.10} \eea where $g$ and $h$ are the open and closed
string coupling constants $(h \sim g^2)$, respectively. Tachyon
condensation with this model was analyzed in \cite{schnabl}.
According to the above adopted approach for construction of the
zeta  string Lagrangian we start with

\bea  {L}' &=& \nn \frac{1}{g^2}\, \sum_{n\geq 1} \frac{n
-1}{n^2}\,\, {\mathcal M}_n (\phi, \theta)  +  \frac{1}{h^2}\,
\sum_{n\geq 1} \frac{n^2 -1}{n^4} \,\, {\mathcal N}_n (\theta) \\
\nn &=&\frac{1}{g^2}\, \sum_{n\geq 1} \Big[ - \frac{1}{2} \phi\,
n^{-\frac{\Box}{2}}\, \phi + \frac{1}{n+1}\,
\theta^{\frac{n(n-1)}{2}}\, (\phi^{n+1} - 1) \Big]\\ &+&
\frac{1}{h^2}\, \sum_{n\geq 1} \Big[- \frac{1}{2} \, \theta\,
n^{-\frac{\Box}{4}}\, \theta + \frac{1}{n^2 +1}\, \theta^{n^2 +
1}\Big]\,,  \label{1.10} \eea where infinite power series are
convergent for $ |\phi| < 1 $ and  $  |\theta| < 1$.

Using again  Riemann's zeta function definition (\ref{1.4}) one
obtains Lagrangian \bea L' &=& \frac{1}{g^2} \Big[- \frac{1}{2}\,
\phi \, \zeta\Big(\frac{\Box}{2} \Big) \phi +  \sum_{n\geq 1}
\frac{1}{n + 1} \, \theta^{\frac{n(n-1)}{2}} \, (\phi^{n+1} -
1)\Big] \nn \\ &+& \, \frac{1}{h^2} \Big[- \frac{1}{2}\, \theta \,
\zeta\Big(\frac{\Box}{4} \Big) \theta +  \sum_{n\geq 1}
\frac{1}{n^2 + 1} \, \theta^{n^2 +1} \Big] \label{1.11}\eea for
the coupled zeta strings $\phi$ and $\theta$, which are open and
closed, respectively.

The equations of motion are

\beq   \zeta \Big( \frac{\Box}{2} \Big)\, \phi =
\frac{1}{(2\pi)^D}\, \int e^{ ixk}\,
\zeta\Big(-\frac{k^2}{2}\Big)\, \tilde{\phi}(k)\,dk = \sum_{n \geq
1} \theta^{\frac{n(n-1)}{2}}\, \phi^n \,,\label{1.12} \eeq

\bea   \zeta \Big( \frac{\Box}{4} \Big)\, \theta &=&
\frac{1}{(2\pi)^D}\, \int e^{ ixk}\,
\zeta\Big(-\frac{k^2}{4}\Big)\, \tilde{\theta}(k)\,dk \nn \\& =&
\sum_{n \geq 1}\, \Big[   \theta^{n^2} + \frac{n (n-1)}{2(n+1)} \,
\theta^{\frac{n(n-1)}{2} -1} \, (\phi^{n+1} -1) \Big]\,,
\label{1.13} \eea and one can easily see  trivial solution $\phi =
\theta = 0$. The weak field approximation  obtains taking only
terms with $n =1$ and related equations of motion are:

\beq  \zeta \Big( \frac{\Box}{2} \Big)\, \phi =  \phi \,, \quad
\zeta \Big( \frac{\Box}{4} \Big)\, \theta = \theta \eeq whose
spatially homogeneous solutions have only trivial solutions, i.e.
$\phi(t) = \theta (t) = 0 $ and there is no mass.

The corresponding potential is

\beq V (\phi, \, \theta) =  \sum_{n \geq 1} \Big[ \frac{1}{g^2} \,
\frac{1}{n+1}\, \theta^{\frac{n(n-1)}{2}} \, ( 1 - \phi^{n+1}) -
\frac{1}{h^2}\, \frac{1}{n^2 +1}\, \theta^{n^2 +1} \Big] \,,
\label{1.14} \eeq for which $V (\phi\,, \theta =0) = 0$ but its
behavior in the region $-1 < \phi <  1\,, \, -1 <\theta < 1$ is
more  complex then potential (\ref{1.8}) and needs an analysis in
detail.

\section{Concluding remarks}
We derived effective field Lagrangians  for description of open
and open-closed bosonic strings, which contain all $p$-adic
Lagrangians. As a result one obtains that an infinite number of
spacetime derivatives and related nonlocality are governed by the
Riemann zeta function. Potentials are nonlinear. The tachyon is
absent in this theory, although  it is contained in the
constitutive $p$-adic Lagrangians. In this model there is no mass.
Energy is bounded from below. $p$-Adic  ingredients can be easily
restored from  the whole Lagrangian using just an inverse
procedure for its construction.

This paper contains only some classical field properties of the
introduced zeta strings. There are still many classical aspects
which need to be investigated. One of them is a systematic study
of the equations of motion. In the quantum sector it is very
desirable to derive and explore scattering amplitudes.

In this short paper we have restricted our attention  to the case
when fields satisfy $|\phi| < 1$ and $|\theta| < 1$, and the
Riemann zeta function $\zeta (s)$ is defined  for ${\mathcal Re}\,
s > 1$.  Analytic continuation of the potentials and the zeta
function should provide more complete insights.

\section*{\large Acknowledgements}
The work on this article was partially supported by the Ministry
of Science and Environmental Protection, Serbia, under contract No
144032D. The author thanks I. Ya. Aref'eva and I. V. Volovich for
useful discussions. This paper was completed during author's stay
in the Steklov Mathematical Institute, Moscow.

 \end{document}